\documentclass[reprint,
superscriptaddress,
twocolumn,
 amsmath,amssymb,
 aps,
pra,
]{revtex4-2}

\usepackage{graphicx}
\usepackage{epstopdf}
\usepackage{epsfig}
\usepackage{xcolor}
\usepackage{subcaption}
\usepackage{dcolumn}
\usepackage{bm}
\usepackage{bbold}

\usepackage{caption}
\usepackage{mathrsfs}
\makeatletter
\renewcommand\@makecaption[2]{%
  \par
  \vskip\abovecaptionskip
  \begingroup
   \small\rmfamily
    \begingroup
     \samepage
     \flushing
     \let\footnote\@footnotemark@gobble
     \@make@capt@title{#1}{#2}\par
    \endgroup
  \endgroup
  \vskip\belowcaptionskip
}
\makeatother

\begin{document}

\preprint{APS/123-QED}

\title{Diffusiophoresis-Enhanced Turing Patterns}

\author{Benjamin M. Alessio}
\affiliation{Department of Chemical and Biological Engineering, University of Colorado, Boulder}
\author{Ankur Gupta}
\email{Corresponding author:ankur.gupta@colorado.edu}
\affiliation{Department of Chemical and Biological Engineering, University of Colorado, Boulder}


\begin{abstract}
Turing patterns are fundamental in biophysics, emerging from short-range activation and long-range inhibition processes. However, their paradigm is based on diffusive transport processes, which yields Turing patters that are less sharp than the ones observed in nature. A complete physical description of why the Turing patterns observed in nature are significantly sharper than state-of-the-art models remains unknown. Here, we propose a novel solution to this phenomenon by investigating the role of diffusiophoresis in Turing patterns. The inclusion of diffusiophoresis enables one to generate patterns of colloidal particles with significantly finer length scales than the accompanying chemical patterns. Further, diffusiophoresis enables a robust degree of control that closely mimics natural patterns observed in species like the Ornate Boxfish and the Jewel Moray Eel. We present a scaling analysis indicating that chromatophores, ubiquitous in biological pattern formation, are likely diffusiophoretic, and that colloidal Péclet number controls the pattern enhancement. This discovery suggests important features of biological pattern formation can be explained with a universal mechanism that is quantified straightforwardly from the fundamental physics of colloids and inspires future exploration of adaptive materials, lab-on-a-chip devices, and tumorigenesis.

\end{abstract}

\maketitle
In his seminal paper, Alan Turing proposed that reaction-diffusion instabilities could give rise to pattern formation in biological systems, a phenomenon now known as Turing patterns~\cite{turing1952chemical}. Subsequent research has provided experimental evidence for the existence of Turing patterns in a variety of biological contexts, including zebrafish embryogenesis~\cite{muller2012differential}, hair follicle spacing~\cite{sick2006wnt}, emulsion-based chemical cells~\cite{tompkins2014testing},  patterns found on zebrafish through cell-cell interactions~\cite{nakamasu2009interactions}, and the development of fingers from early limb-buds~\cite{raspopovic2014digit}. Theoretical advances have led to the development of modifications to Turing's original model~\cite{koch1994biological}, including the widely-used Gierer-Meinhardt~\cite{gierer1972theory}, Brusselator~\cite{prigogine1967symmetry, prigogine1968symmetry} and cell-cell interaction~\cite{nakamasu2009interactions} models. These models and their derivatives have been used to reproduce a wide range of patterns found in nature~\cite{vittadello2021turing}, from the wave-like patterns on marine angelfish~\cite{kondo1995reaction}, to the stripe-like patterns on zebrafish~\cite{nakamasu2009interactions,kondo2010reaction}, and even patterns on sea shells~\cite{meinhardt2009algorithmic}. However, these models typically rely on diffusive transport mechanisms to generate concentration gradients and have thus far have neglected the role of convective transport. We propose that, in the context of Turing patterns, cells convect and steepen using diffusiophoresis; this mechanism promotes the color sharpening that is observed in nature.

\par{} Previous attempts to replicate the sharp gradients observed in natural Turing patterns have prescribed ad-hoc mechanisms including switching systems~\cite{koch1994biological} and reaction pathway coupling~\cite{sanderson2006advanced}. However, a complete physical justification for such processes remains elusive. A more natural mechanism to obtain sharp gradients in concentration comes from operating in the regime of high P\'{e}clet numbers, i.e., conditions when the convective transport dominates the  diffusive transport. This regime might appear to be difficult to obtain in the absence of fluid flow. However, such a regime is commonly observed in microfluidic experiments with colloidal particles via the process of diffusiophoresis, i.e., transport of colloidal particles in response to solute concentration gradients \cite{abecassis2008boosting, theurkauff2012dynamic, banerjee2019long, shin2016size, kar2015enhanced, singh2020reversible, banerjee2016soluto, shi2016diffusiophoretic, wilson2020diffusiophoresis, akdeniz2023diffusiophoresis, gupta2020diffusiophoresis, alessio2022diffusioosmosis, alessio2021diffusiophoresis}. In fact, the key feature observed in diffusiophoretic systems is the banding of colloidal particles that creates region of sharp colloidal concentration gradients~\cite{abecassis2008boosting, raj2023two, shin2016size, shi2016diffusiophoretic, wilson2020diffusiophoresis, alessio2021diffusiophoresis, alessio2022diffusioosmosis}. The diffusiophoretic transport of colloids, which has been seen to occur with a P\'{e}clet number $\textrm{Pe} = O(10)-O(10^3)$~\cite{abecassis2008boosting, shi2016diffusiophoretic,  alessio2021diffusiophoresis, alessio2022diffusioosmosis} (the ratio of particle convective flux to diffusive flux), creates these banded structures that can be further enhanced in the presence of acid-base reactions~\cite{banerjee2019long}. We note that chemotaxis can be qualitatively similar to diffusiophoresis and has been proposed in the past~\cite{painter1999stripe} as an explanation for subtle features arising from time-dependent processes in the development of angelfish. Such an effect has, however, been neglected in more recent state-of-the-art explanations of chromatophore pattern formation~\cite{yamaguchi2007pattern,nakamasu2009interactions}. As we will argue, chromatophores are likely diffusiophoretic. Thus, we can recover key features by incorporating colloid physics into models of biological pattern formation models.

\par{} Over the past decade, several studies have concluded that diffusiophoresis is a key process that was previously overlooked. For instance, Florea et al. \cite{florea2014long} elucidated, both theoretically and experimentally, that diffusiophoresis is responsible for creating an exclusion zone near charged surfaces and in fact stated that diffusiophoresis is ``likely to play an important, yet unexplored role" in biological processes. Similarly, Shin et al. \cite{shin2018cleaning} underscored the role of diffusiophoresis in removing contaminants from porous, fibrous materials in the presence of surfactant concentration gradients. Therefore, we surmise that diffusiophoresis of chromatophores could play an important role in the process of biological pattern formation. To the best of our knowledge, this effect has not been investigated.

\par{} Some of the most readily observable Turing patterns are animal skin patterns. Our central argument is that during biological pattern formation, chromatophores, which are specialized pigment cells known to control the coloration pattern on fish~\cite{fujii2000regulation}, respond diffusiophoretically to physiological reactions. There are different types of chromatophores, including melanophores, xanthophores, erythrophores, iridophores and leucophores~\cite{williams2019dynamic}, which are organized in multiple layers and can interact with each other to create complex, even regenerative or adaptive patterns~\cite{bagnara1968dermal,marshall2019colours}. We postulate that chromatophores are diffusiophoretic because of three primary reasons. First, the size of a typical chromatophore can range from 1-30 microns, which is typically the range where colloidal particles are diffusiophoretic~\cite{velegol2016origins, shim2022diffusiophoresis}. Second, recent experimental results suggest that chromatophores are charged and interact with the surrounding medium~\cite{liu2023rotary}. Finally, the biological reactions in the surrounding medium can create concentration gradients of small molecules that will interact with the chromatophore; multi-component physiological reactions have been known to control chromatophore aggregation and dispersion~\cite{kawauchi1983characterization}. We also emphasize that diffusiophoresis of chromatophore-coated particles has been experimentally reported in literature~\cite{liu2023rotary}. In a biological setting, we expect that particle motion is made complex by effects such as cell-cell adhesion and resistance of the porous membrane. Regardless, such effects do not impact the essential physics leading to steepening.

\par
\begin{figure}
\centering
\includegraphics[width=0.45\textwidth]{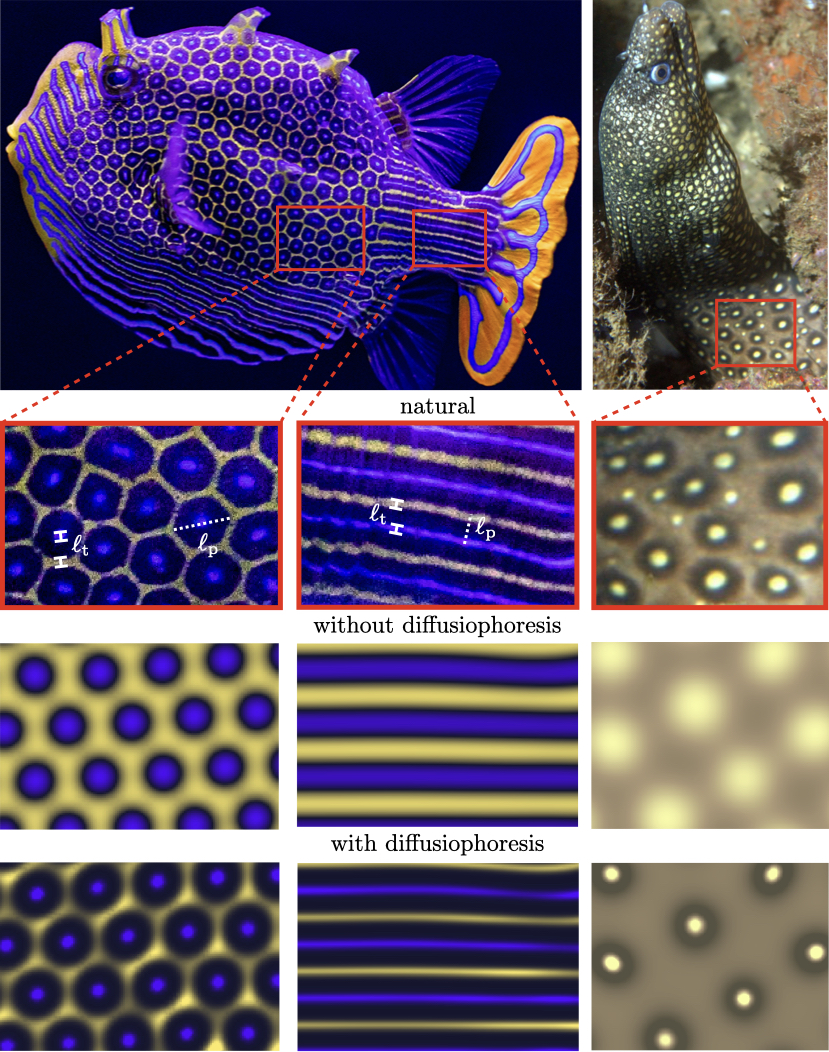}
\caption{ \textbf{Comparison of natural patterns in fish simulations.} (Left) A male Ornate Boxfish (\textit{Aracana ornata}) has intricate hexagon and stripe patterns on its skin. Reaction-diffusion models can capture the nature of the pattern but are unable to reproduce the sharpness of the color gradients. Instead, diffusiophoresis-enhanced reaction-diffusion shows a striking resemblance to the natural patterns when compared to simulations from our model. Photo courtesy of the Birch Aquarium at the Scripps Institution of Oceanography. White line segments and dashed lines indicate the edge thickness, i.e., $\ell_{\textrm{t}}$ and pattern sizes $\ell_{\textrm{p}}$, respectively. (Right) A Jewel Moray Eel (\textit{Muraena lentiginosa}). The double spot pattern cannot be reproduced from a two-component reaction-diffusion dynamics alone, but is easily simulated as a diffusiophoretic process. Image by craigjhowe, used under CC BY-NC 4.0 license. Source: iNaturalist observation 37252428. }\label{fish}
\end{figure}

{} The minimum mathematical model that qualitatively reproduces the features of biological patterns consists of two reactive-diffusive species that interact with each other through a nonlinear reaction mechanism~\cite{turing1952chemical, kondo1995reaction, kondo2010reaction, gierer1972theory, prigogine1967symmetry, prigogine1968symmetry}.  The most natural interpretation in the minimum mathematical model is to assume that these two species are chromatophores. We compare the natural patterns observed in Ornate Boxfish (\textit{Aracana Ornata}) with the Brusselator model (see Methods for details) assuming a blue and a yellow chromatophore to be our reactive species. As observed in figure~\ref{fish}, the Brusselator model is able to reproduce the patterns of hexagon and stripes seen in the Boxfish, though it is unable to capture the sharpness of color gradients observed in the fish. Besides the size of the fish, in the natural fish patterns, there are two distinct length scales (see Methods). For instance, for the hexagon pattern, there is a clear length scale separation between the edge-thickness, $\ell_\textrm{t}$, and the length of the hexagon pattern, $\ell_\textrm{p}$. Similarly, for the stripe pattern, there is a distinction between the the thickness of the stripes $\ell_\textrm{t}$ and the separation between the stripes $\ell_\textrm{p}$. The reaction-diffusion models have an inherent limitation that $\ell_\textrm{t}=\ell_\textrm{p}$, leading to a diffuse pattern formation. Accordingly, they are unable to recover double spot patterns such as those seen on the Jewel Moray Eel (\textit{Muraena lentiginosa}) unless multiple nonlinear reaction mechanisms are invoked, i.e., two or more Brusselators are coupled~\cite{sanderson2006advanced}, which still yield a diffuse pattern unlike the ones observed in nature. Clearly, we must look for mechanisms beyond the nonlinear reaction-diffusion interactions.   
\par{} While experiments~\cite{yamaguchi2007pattern,nakamasu2009interactions} have soundly implicated chromatophores in pattern-formation in zebrafish skin, recent evidence~\cite{nakamasu2009interactions} suggests that a third species, which is more diffusive and is thus a small molecule, is required for a more complete picture of the process. This invites the possibility of robust control of separate length scales, where $\ell_\textrm{p}$ is associated with the molecular substance(s) and $\ell_\textrm{t}$ is associated with the chromatophores. Furthermore, prior studies on two-component models utilized species diffusion coefficients $O(10^{-10})-O(10^{-9})$~m$^2/$s to obtain patterns that resemble the ones observed in nature~\cite{kondo1995reaction}. These diffusivity values resemble small molecules instead of microparticles, which have diffusivities $O(10^{-13})$~m$^2/$s. Accordingly, we assert that there are two broad categories of species present in the system: molecular-scale solute species (i.e. morphogens or long-range mediators) and microscopic-scale colloidal species (i.e. chromatophores). This is supported by prior experiments~\cite{kawauchi1983characterization}.

\par{}  The transport equations for two-component chromatophore models are primarily dependent on the Damk\"{o}hler number of the colloids $\textrm{Da}_N$, which is the ratio of the diffusion time scale to the reaction time scale, where an increase in $\textrm{Da}_N$ corresponds to a decrease in $\ell_\text{p}$. However, in our proposed model that builds on the recent experimental findings~\cite{nakamasu2009interactions}, the reactions are driven by solutes and the chromatophores respond to the concentration gradients of the solutes diffusiophoretically; see Methods for derivation. In this proposed setup, the two dimensionless groups that control the behavior of the patterns are the Damk\"{o}hler number of the solute, $\textrm{Da}_C$, and the P\'{e}clet number of the colloids, $\textrm{Pe}$. Typically, $\textrm{Pe} = O(10) - O(10^3)$ \cite{abecassis2008boosting, wilson2020diffusiophoresis, gupta2020diffusiophoresis}. Physically, we propose that transport of the solute molecules sets only the pattern type and size $\ell_\textrm{p}$, whereas the diffusiophoretic transport of colloids dictates only the thickness $\ell_\textrm{t}$.

\par
\begin{figure}
\centering
\includegraphics[width=0.45\textwidth]{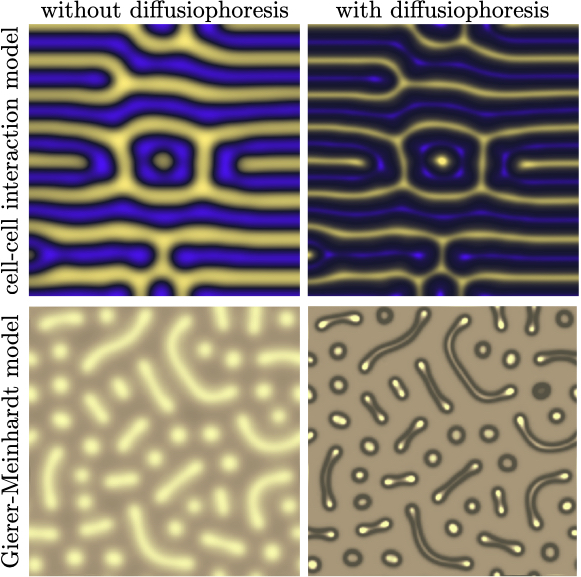}
\caption{\textbf{Diffusiophoretic enhancement for various reaction-diffusion mechanisms}. (Top row) Cell-cell interaction model~\cite{nakamasu2009interactions}. (Bottom row) Gierer-Meinhardt model~\cite{gierer1972theory}. For both rows, the left panel shows a stable pattern arising from purely reactive-diffusive mechanisms, and the right panel shows the steepening effect of diffusiophoresis. }\label{otherModels}
\end{figure}

{} Our proposed model is able to recover to an considerable degree the natural patterns of hexagons and stripes observed in the Ornate Boxfish as well as the double-spot pattern in the Jewel Moray Eel using the Brusselator model with diffusiophoresis; see figure~\ref{fish} and Supplementary Videos 1 and 2. While the reaction-diffusion model without diffusiophoresis produces qualitative features of the chromatophore patterns, by including diffusiophoresis we capture the steepening effect. This feature is universal and not exclusive to the Brusselator model. In figure~\ref{otherModels} and Supplementary Videos 3 and 4, we perform the same comparison between reaction-diffusion models with and without diffusiophoresis for both the Gierer-Meinhardt model~\cite{gierer1972theory} and cell-cell interaction model~\cite{nakamasu2009interactions} (see Methods: Numerical simulations for details). In these models, just as with the Brusselator in figure 1, the sharpening effect of diffusiophoresis is apparent both in the finer length scales and the greater magnitude of depletion away from the hotspots (i.e., the color of the two phases is more distinct with the inclusion of diffusiophoresis). Because of its relative abundance and depth of analytical theory in the literature, in this paper we center our analysis on the Brusselator model, but the fundamental physics at play is agnostic to the model.
\begin{figure*}
\centering
\includegraphics[width=\textwidth]{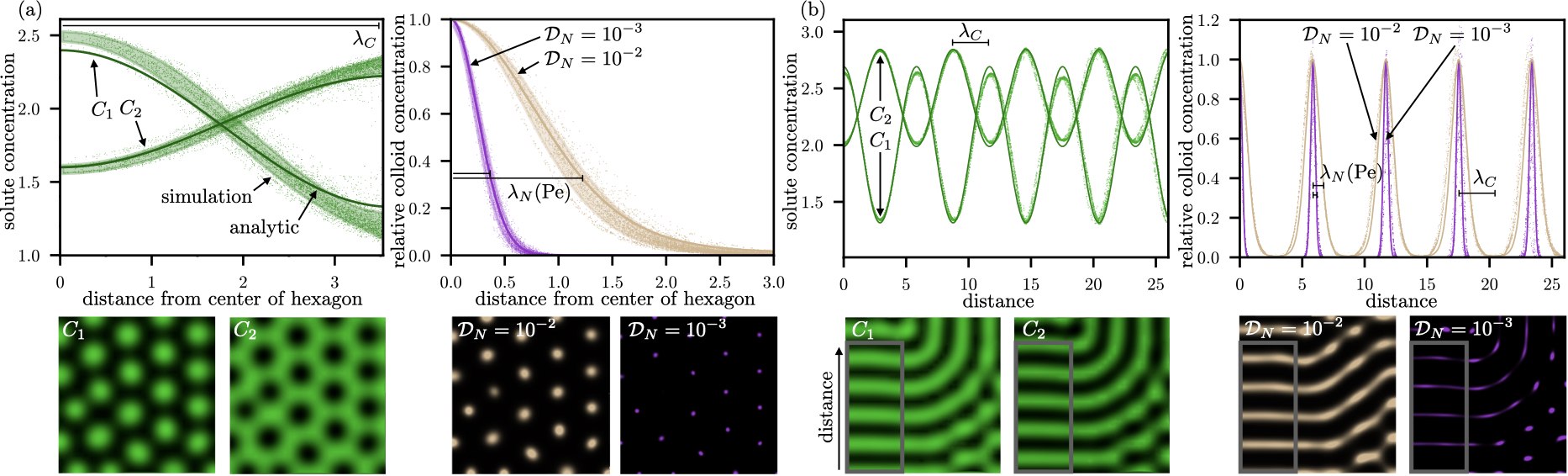}
\caption{\textbf{Controlling the edge-thickness of Brusselator patterns}. (a) Comparison of analytically derived and numerically computed forms of (left) solutes and (right) colloidal concentration profiles, averaged radially over all hexagons in the frame. The solute concentrations are simulated with parameters $\mu=0.05$, $A=1.5$, and $\mathcal{D}_{C_2}=4$.  The colloidal concentrations correspond to two different simulations $\mathcal{D}_N=10^{-2}$ and $\mathcal{D}_N=10^{-3}$, migrating with $M_1=M_2=0.1$ for both cases. Solute and colloidal length scales are indicated with line segments. The shading in all curves represents one standard deviation of the numerical data points. (b) Comparison of analytically derived and numerically computed forms of (left) solutes and (right) colloidal concentration profiles with vertical distance, averaged across the stripes in the gray boxes. The solute concentrations are simulated with parameters $\mu=0.04$, $A=2$, and $\mathcal{D}_{C_2}=3$. The colloidal concentrations correspond to two different simulations $\mathcal{D}_N=10^{-2}$ and $\mathcal{D}_N=10^{-3}$, migrating with $M_1=M_2=0.1$ for both cases. Solute and colloidal length scales are indicated with line segments.}\label{hexagonsStripes}
\end{figure*}

\par{}To further emphasize the distinction between the proposed and the reaction-diffusion mechanisms, in figure~\ref{Da}, we compare the hexagon pattern for different values of $\textrm{Da}_N$ in the reaction-diffusion model and different values of Pe in the proposed model. As is clear from the results, increasing $\textrm{Da}_N$ simply decreases $\ell_{\textrm{p}}$ without changing the relative thickness of concentration gradients, i.e., $\frac{\ell_\textrm{t}}{\ell_\textrm{p}}$ is constant. In contrast, an increase in Pe does not modify $\ell_\textrm{p}$ but reduces $\ell_\textrm{t}$, and better resembles the patterns observed in nature. In our simulations, we found strongest qualitative agreement with natural patterns when setting Pe in the range $10-50$. This is consistent with the effective Pe number observed in recent theoretical and experimental findings \cite{alessio2021diffusiophoresis, alessio2022diffusioosmosis}.  We acknowledge that we neglect inter-particle interactions and invoke the assumption of dilute suspension, which in some limits may cause our simulations to under-predict values of $\ell_\textrm{t}$. 

\par{} Next, we focus on quantifying the pattern formation of chromatophores via diffusiophoresis. For simplicity, we focus on the Brusselator model for solute reactions (figure~\ref{schematic} left panel), though our analysis is readily extended to other reaction-diffusion frameworks \cite{kondo2010reaction}. In the Brusselator model~\cite{prigogine1967symmetry,prigogine1968symmetry}, two abundant reactants with concentrations $A$ and $B$ produce solutes with concentrations $C_1$ and $C_2$, where concentrations are scaled by $c^*$. An initially homogeneous state of $C_1$ and $C_2$ can spontaneously come into formation when perturbations of permitted wavelengths grow into patterns such as hexagons or stripes; the resulting gradients $\nabla C_1$ and $\nabla C_2$ generate the diffusiophoretic velocity $V_{\text{DP}_j} = M_{j1} \nabla C_1 + M_{j2} \nabla C_2$ of the colloid (i.e. chromatophore) for diffusiophoretic mobility $M_i$ corresponding to $C_i$, where $M_{ji}$ is the dimensionless diffusiophoretic mobility that induces the velocity in the $j^{\textrm{th}}$ colloid due to the $i^{\textrm{th}}$ solute. The results in figure~\ref{schematic} highlight that two different diffusiophoretic colloids with different mobilities, denoted with white and red dots, mirror the pattern of the solute to form hexagons and stripes, but produce sharper and more well-defined features than the solute alone. Only concentration hotspots of $C_1$ are shown (in green) since the concentration distribution of $C_2$ has the same spatial structure as $C_1$. Here, we further reduce the parameter space by focusing on a system with one colloid only and therefore we drop the $j^{\textrm{th}}$ subscript in $M_{ji}$ and refer to it as $M_{i}$.

\par{} By employing the framework of amplitude equations, we analytically predict solute and relative colloid concentrations for different patterns produced with the Brusselator model, including an analytical expression for the steepening ratio $\frac{\lambda_N}{\lambda_C}$; see Methods for details. Theoretical investigations of the Brusselator model~\cite{pena2001stability} have revealed a complex parameter space where different Turing patterns form. The perturbations are sinusoidal in nature and have amplitudes that obey complex partial differential equations that encode the stability and form of these structures. It is worth noting that these equations are not deterministic, and merely describe the shape and stability of all possible structures~\cite{vittadello2021turing} such that in a simulation, any stable pattern can be forced, and in experiments, we expect to the fastest-growing wavelength to dominate. 

\par{} One of the dominant patterns in the Brusselator model is hexagons. Simulating stable hexagon patterns, in figure~\ref{hexagonsStripes}(a) we plot the dependence of solute (left) and colloidal (right) concentrations with distance away from the center of the hexagon (averaged over all hexagons in the simulation box). The other dominant pattern in the Brusselator model is stripes, which we investigate in figure~\ref{hexagonsStripes}(b). The solute (left) and colloidal (right) concentrations here are plotted along the periodic direction.

There are two length scales in each of the patterns described in figure~\ref{hexagonsStripes}. In hexagons, it is the size of the hexagons and the edge-thickness of the hexagons. For stripes, it is the spacing between the stripes and the thickness of the stripes. The size of the hexagons and the spacing between the stripes is set by $\lambda_C$, i.e., the wavelength of the solute patterns, and the thickness of hexagons and stripes is set by $\lambda_N$. Crucially, the ratio of $\frac{\lambda_N}{\lambda_C}$ is set by the P\'{e}clet number of the colloid for both the patterns. To explore this effect in more detail, we mathematically derive (see Methods for derivation) the P\'{e}clet number as
\begin{equation}
    \text{Pe} = \bigg\rvert\frac{\alpha\left(M_1-M_2\frac{\eta(1+A\eta)}{A}\right)}{\mathcal{D}_N}\bigg\rvert,
    \label{eqn:peclet dimensional}
\end{equation}
where $\alpha$ is the amplitude of the perturbations to $C_1$ and $\eta = \mathcal{D}_{C_2}^{-1/2}$. We note that equation~(\ref{eqn:peclet dimensional}) is valid for both hexagon and stripe patterns. In the Brusselator model, the non-dimensional solute concentration local to hexagon or stripe structures as a function of distance $\mathscr R$ from the local maximum is given by
\begin{equation}
    C_i(\mathscr R) = \sigma_i\alpha\cos\left(2\pi\frac{\mathscr R}{\lambda_C}\right) + \text{const},
    \label{eqn:Ci_analytic}
\end{equation}
where we note that the one-dimensional form of equation~(\ref{eqn:Ci_analytic}) is exact in the direction of maximal gradient for stripes and approximate for radially-averaged hexagons. From the gradient of equation~(\ref{eqn:Ci_analytic}), we calculate the diffusiophoretic velocity
\begin{equation}
    V_\text{DP}=-\frac{2\pi}{\lambda_C} \alpha\left(M_1-\frac{\eta(1+A\eta)}A M_2\right)\sin\left(2\pi\frac{\mathscr R}{\lambda_C}\right)
    \label{eqn:Vdp_bruss}
\end{equation}
and the form of the steady-state colloid concentration
\begin{equation}
    N(\mathscr R) \propto \exp\left( \text{Pe}\cos\left(2\pi\frac{\mathscr R}{\lambda_C}\right)\right).
    \label{eqn:N_analytic}
\end{equation}
We define the colloid length-scale $\lambda_N$ as the exponential decay distance, giving the analytical expression
\begin{equation}
    \cos\left(2\pi\frac{\lambda_N}{\lambda_C}\right) = 1 - \frac1{\text{Pe}};
    \label{eqn:masterCurve}
\end{equation}
see Methods: Analytical model for derivations. In figure~\ref{hexagonsStripes}, we show that our analytical model is in good quantitative agreement with simulation data extracted from a large number of both hexagons and stripes for solute and colloidal concentrations. We note that the analytical model has no discontinuities, and the colloid hotspots are predicted on a continuous and periodic curve.
\par
\begin{figure}
\centering
\includegraphics[width=0.49\textwidth]{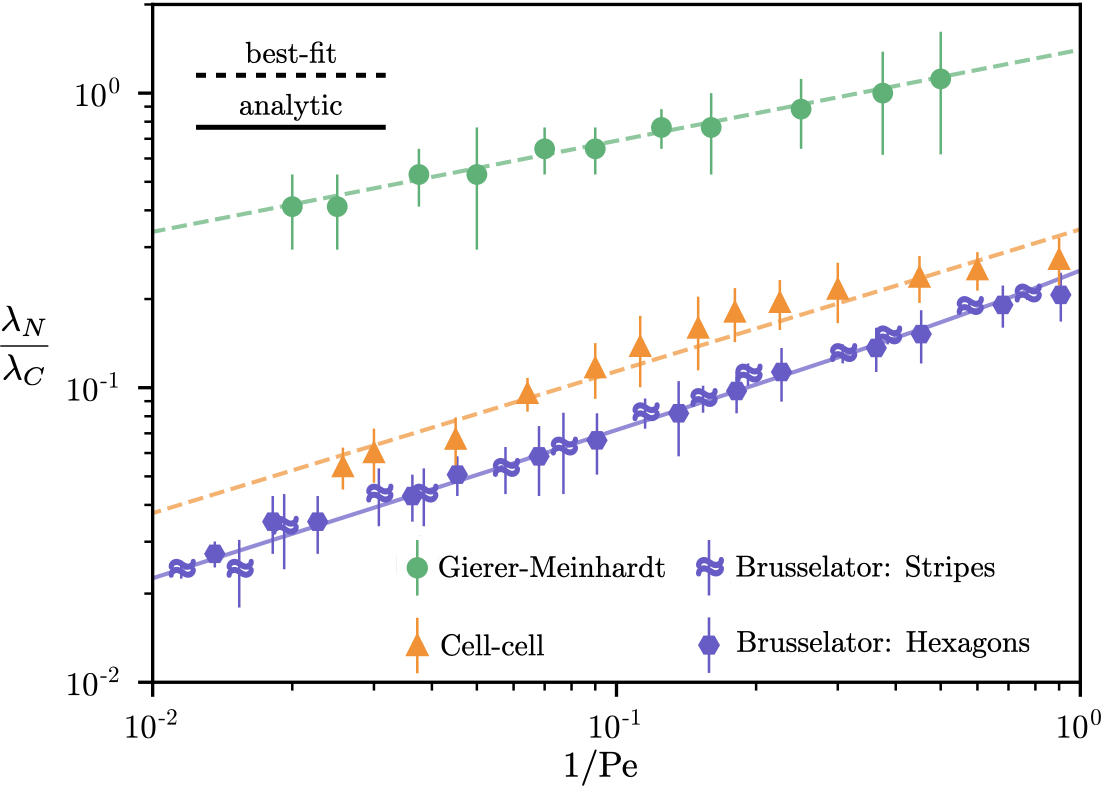}
\caption{\textbf{Master curve of diffusiophoretic-enhanced Turing patterns}. Comparison of simulation results with analytical predictions for the colloid hotspot length scale $\lambda_N$ across a wide range of Péclet numbers. For the Brusselator model,  hotspot lengthscales over both hexagon and stripe patterns are compared to an analytical curve with no fitting parameters. The same model parameters used in figure~\ref{hexagonsStripes} are applied. For the Gierer-Meinhardt and cell-cell interaction models, data points are compared to a best fit power law (see Methods: Numerical simulations for details). Error bars represent one standard deviation in computations.}\label{masterCurve}
\end{figure}

{} It is clear from equation~(\ref{eqn:masterCurve}) that $\frac{\lambda_N}{\lambda_C}$ can vary significantly through changes in Pe, revealing a control mechanism for the sharpness of Turing patterns involving colloidal particles. We further elucidate this control mechanism in figure~\ref{masterCurve} by comparing numerical predictions of $\frac{\lambda_N}{\lambda_C}$ vs. $\textrm{Pe}$ to the analytical theory for the Brusselator model (equation~\ref{eqn:masterCurve}). We obtain an approximately linear relationship between the two variables, consistent with our theoretical predictions, and show a collapse for two orders of magnitude of $\textrm{Pe}$. In the limit of low Pe, the steepening effect is diminished and the colloid length scale approaches the solute length scale. If Pe is exactly zero, however, there would be no mechanism for colloidal formation and thus $\frac{\lambda_N}{\lambda_C}$ would be undefined. 

\par{} In addition to the Brusselator model, we include in figure~\ref{masterCurve} simulation data points for the Gierer-Meinhardt and the cell-cell interaction models. In the absence of an analytical prediction for the form of the solute field perturbations in these models, we estimate $\textrm{Pe}=\frac{M}{\mathcal{D}_N}$, noting that we observe perturbation amplitudes of $\mathcal{O}(1)$ in our simulations. A more accurate representation of Péclet number would be the general form of equation~(\ref{eqn:peclet dimensional}), $\textrm{Pe}=\sum_i\alpha_i\frac{M_i}{\mathcal{D}_N}$, where the sum is across each solute species and the coefficients $\alpha_i$ are dependent upon the specific solute reaction-diffusion model. However, the analytical theory necessary to obtain such coefficients does not exist, to the best of our knowledge, for the Gierer-Meinhardt \cite{gierer1972theory} and cell-cell \cite{nakamasu2009interactions} interaction models. Nonetheless, the data points from our simulations of these models are well-represented by a best-fit line, suggesting the collapse to a master curve of $\frac{\lambda_N}{\lambda_C}$~vs.~Pe for models other than the Brusselator.

\par{} The findings reported in this article open up numerous possibilities for future research and applications. For instance, further exploration of biological systems could reveal more instances where diffusiophoresis contributes to pattern formation such as embryo morphogenesis, which could subsequently lead to a better understanding of how these patterns emerge, evolve, and adapt in different species. In recent years, an embryonic framework has gained attention for understanding cancer ecosystems as chaotic expressions of the complex interactions between cellular and chemical components~\cite{huang2009cancer}, and Turing patterns are theorized to play an important role in tumorigenesis~\cite{uthamacumaran2020cancer}. Unveiling diffusiophoresis as a mechanism in cellular organization processes can be crucial to future cancer research. Additionally, a more comprehensive understanding of the interaction between different species of chromatophores and solutes might be achieved by exploring various reaction rates, diffusivities, and mobilities. Going forward, research might consider complex colloidal interactions~\cite{riva2022solution}, dense packing effects~\cite{zaccone2022explicit}, anisotropy~\cite{shin2022shape}, particle self-propulsion~\cite{frankel2014dynamics}, and hydrodynamic interactions, which could reveal important control mechanisms of pattern formation in various biological systems. Experimental studies can observe the dynamics of the formation of Turing patterns with colloidal particles in controlled conditions. Beyond biophysics, understanding the role of diffusiophoresis in pattern formation has potential applications in engineering and materials science. By harnessing these mechanisms, researchers could develop novel methods for fabricating materials with precise micro-scale patterns. These materials could find applications in areas such as photonics, lab-on-a-chip devices, and biotechnology. In conclusion, our findings on the role of diffusiophoresis in pattern formation provide a foundation for future research and have the potential to impact a wide range of fields.

\nocite{}
\bibliographystyle{naturemag}
\bibliography{main}

\section*{Methods}

\subsection*{Derivation of transport equations for proposed model}
We consider a dilute mixture of two different types of species: small molecule solutes and micro-scale colloids. The concentration of small molecules is denoted by $c_i$, where $i \in [1,S_C]$ such that $S_C$ is the total number of solute molecule species. The concentration of colloids are given by $n_j$, where $j \in [1,S_N]$ such that $S_N$ is the total number of microparticle species. The transport equation of solutes and colloids are given by 
\begin{subequations}
\begin{equation}
    \frac{\partial c_i}{\partial t} = D_{c_i} \tilde{\nabla}^2 c_i + R_{c_i},
\end{equation}\begin{equation}\label{eqn:dimensional n}
    \frac{\partial n_j}{\partial t} + \tilde{\nabla} \cdot (\mathbf{v}_{\textrm{DP}j} n_j) = D_{n_j} \tilde{\nabla}^2 n_j, 
\end{equation}
\end{subequations}
where $t$ is time, and $D_{c_i}$ and $D_{n_j}$ are the diffusivities of the $i^{\textrm{th}}$ solute and the $j^{\textrm{th}}$ colloid, respectively. $R_{c_i}$ is the production rate of the $j^{\textrm{th}}$ colloid. The term $\mathbf{v}_{\textrm{DPj}}=\sum_i m_{ij}\tilde\nabla c_i$ represents the induced diffusiophoretic velocity of the $j^{\textrm{th}}$ colloid due to the concentration gradients of solutes, proportional to diffusiophoretic mobility coefficient $m_{ij}$. For simplicity, we use the non-electrolyte form of $\mathbf v_\text{DP}$~\cite{raj2023two}, but more complicated forms are well-described in the literature~\cite{gupta2019diffusiophoretic,alessio2021diffusiophoresis}. There is no background fluid flow, although realistic experimental settings are expected to have a strong influence of background flow originating from the solute-geometry interactions~\cite{alessio2021diffusiophoresis,alessio2022diffusioosmosis}. We note that the sign of the divergence in equation~(\ref{eqn:dimensional n}) affects the shape of the steepening pattern; see Methods: Analytical model. Before proceeding further, we introduce non-dimensional quantities where $C_i=\frac{c_i}{c^*}$, $N_i = \frac{n_i}{n^*}$, $T=\frac{t D_{c_1}}{\ell^2}$, $\mathcal{D}_{C_i} = \frac{D_{c_i}}{D_{c_1}}$, $\mathcal{D}_{N_i} = \frac{D_{n_i}}{D_{c_1}}$, $\mathcal{R}_{c_i} = \frac{R_{c_i} D_{c_i}}{c^* \ell^2}$,  $\mathbf{V}_{\textrm{DP}j} = \frac{\mathbf{v}_{\textrm{DP}j} \ell}{D_{c_1}}$, and $\nabla = \ell \tilde{\nabla}$, where $c^*$ and $n^*$ are reference concentrations for solute and colloids, respectively, $\ell$ is a reference length scale, and $D_{c_1}$ is the diffusivity of the first solute. These substitutions yield the equations 
\begin{subequations}
\begin{equation}
\frac{\partial C_i}{\partial T} = \mathcal{D}_{C_i} \nabla^2 C_i + \mathcal{R}_{c_i},
\label{eqn:dCdt}\end{equation}\begin{equation}
\frac{\partial N_j}{\partial T} + \nabla \cdot (\mathbf{V}_{\textrm{DP}_j} N_j) = \mathcal{D}_{N_j} \nabla^2 N_j.
\label{eqn:dNdt}\end{equation}
\end{subequations}
To induce Turing patterns, we define a four-component model with two solute species and two chromatophore species. We use the Brusselator model for solute species to describe the reaction and production rates, that are given by~\cite{prigogine1967symmetry, prigogine1968symmetry} 
\begin{subequations}
\begin{equation}
\mathcal{R}_{c_1} = \textrm{Da}_C (A - (B+1)C_1 + C_1^2C_2),
\end{equation}\begin{equation}\mathcal{R}_{c_2} = \textrm{Da}_C (BC_1 - C_1^2C_2).
\end{equation}\end{subequations}
$A$ and $B$ are scaled concentrations of excess components (see \cite{pena2001stability} for details), $\textrm{Da}_C = \frac{k \ell^2}{D_{c_1}}$ is Damk\"{o}hler number for the solutes and $k$ is the first-order reaction constant.  We define diffusiophoretic velocities for the two colloids as follows 
\begin{subequations}
\begin{equation}
    \mathbf{V}_{DP_1} = M_{11} \nabla C_1 +  M_{21} \nabla C_2,
\end{equation}\begin{equation}
    \mathbf{V}_{DP_2} = M_{12} \nabla C_1 +  M_{22} \nabla C_2,
\end{equation}
\end{subequations}
where $M_{11}$, $M_{21}$, $M_{12}$ and $M_{22}$ are diffusiophoretic mobility coefficients. We note that for electrolytic diffusiophoresis, the dependence of diffusiophoretic velocity is with the gradient of the log of the concentrations \cite{velegol2016origins, gupta2020diffusiophoresis}. However, since the exact nature of interactions between solute and biological colloids remain unknown \cite{dukhin2010peculiarities}, for simplicity, we employ the relationship for non-electrolytic interactions \cite{raj2023two}. We note that the values of mobility coefficients are related to the $\textrm{Pe}_N$ and are typically at most $O(1)$. We solve Eqs.~(\ref{eqn:dCdt}) and~(\ref{eqn:dNdt}) with initial conditions of unit concentrations and no-flux boundary conditions. The details of numerical simulations are provided in a later section. 
\par{} The two-component reaction diffusion model is identical to the solute equations where $C_1$, $C_2$ are replaced by $N_1$, $N_2$, $D_{C_i}$ is replaced by $D_{N_i}$, $\mathcal{R}_{C_1}$ and $\mathcal{R}_{C_2}$ are changed to $\mathcal{R}_{N_1}$ and $\mathcal{R}_{N_2}$, and $\textrm{Da}_C$ is modified to $\textrm{Da}_N$.   
\subsection*{Scaling arguments}
Our numerical model relies on the following dimensionless groups: $\textrm{Da}_C$ and $\textrm{Pe}$. To estimate $\textrm{Da}_C$, we recall that $\textrm{Da}_C = \frac{k \ell^2}{D_{c_1}}$. For small molecules, $D_{c_1}=O(10^{-10}) - O(10^{-9}) \ \textrm{m}^2/$s. Based on the typical experiments previously reported on fishes~\cite{kondo1995reaction, nakamasu2009interactions}, it appears that $k=O(10^{-7}) - O(10^{-6})~\textrm{s}^{-1}$. We anticipate this number can significantly change for different kinds of fishes, and thus this choice is the best estimate based on prior literature values. To determine an appropriate value of $\ell$, we note that typical biological patterns have three main length scales. The simplest choice is to pick the size of an organism, that for fishes, is typically $ \ell_{\textrm{b}} = O(10^{-2})-O(10^{-1})~\textrm{m}$. The second choice is the length scale of a repeating pattern (such as size of hexagon or the spacing between the stripes), that tends to be one order of magnitude smaller than the body size, or $ \ell_{\textrm{p}} =  O(10^{-3})-O(10^{-2})~\textrm{m}$. Finally, the third choice is the thickness of these patterns (see the thickness of the blue and yellow patterns on the fish in figure~\ref{fish}), that, at least for fishes, appears to be one order of magnitude smaller than $\ell_{\textrm{p}}$, or  $\ell_{\textrm{t}}= O(10^{-4})-O(10^{-3})~\textrm{m}$. Out of these three choices, we postulate that $\ell_{\textrm{p}}$ is the most appropriate choice since $Da_C=O(1)$. This emphasizes the fact that presence of small molecules is a crucial element for pattern formation. However, they only set the pattern size $\ell_{\textrm{p}}$, and not the pattern thickness $\ell_{\textrm{t}}$, which we discuss next.
\par{} To estimate Pe, it is straightforward to see from equation~(\ref{eqn:peclet dimensional}) that $\textrm{Pe} = O\left( \frac{M}{\mathcal{D}_N} \right)$. As per prior literature values, $M \lesssim O(1)$ \cite{gupta2020diffusiophoresis}, and thus $\textrm{Pe} \lesssim O\left( \frac{1}{\mathcal{D}_N} \right)$. For a typical colloid, thus, $\textrm{Pe} = O(10) - O(10^3)$. We note that recent findings have suggested that $\mathcal{D}_N$ is higher that the native diffusivity of the colloid due to dispersion, indicating a lower $\textrm{Pe}$ than expected~\cite{alessio2022diffusioosmosis}. 

\subsection*{Analytical model}\label{sec:analytical model}
We restrict our analysis to chemical patterns exhibiting only critical wavelengths $\lambda_C=2\pi/\sqrt{A\eta}$ (where $\eta\equiv\sqrt{\mathcal{D}_{C_2}}$), noting that this allows for a rich phase space while keeping our model simple~\cite{pena2001stability}. We consider solute concentrations that are one-dimensional in distance $\mathscr R$. These are exact in the direction of maximal gradient of stripe patterns and approximate for radially-averaged hexagons. Setting $\mathscr R=0$ to be the location of the extremum, $\sigma_1=1$, $\sigma_2=-\eta(1+A\eta)/A$, and noting the coefficient $\alpha$ depends on the pattern, we have equation~(\ref{eqn:Ci_analytic}). We define supercriticality parameter $\mu\equiv (B-1-A\eta^2)/(1+A\eta^2)$ and adapt the coefficients from Peña and Pérez-García~\cite{pena2001stability}:
\begin{subequations}
\begin{equation}
    \alpha_\text{stripes} = \text{sign}(f)\sqrt{\mu/g},
\end{equation}
\begin{equation}
    \alpha_\text{hexagons} = 3\frac{f + \text{sign}(f)\sqrt{f^2 + 4\mu(g + 2h)}}{2(g + 2h)},
\end{equation}
\begin{equation}
    f = 2\frac{1-A\eta}{A(1+A\eta)} + \frac2A\mu,
\end{equation}
\begin{equation}
    g = \frac{-8 + 38A\eta + 5A^2\eta^2 - 8A^3\eta^3}{9A^3\eta(1+A\eta)},
\end{equation}
\begin{equation}
    h = \frac{-3 + 5A\eta + 7A^2\eta^2 - 3A^3\eta^3}{A^3\eta(1+A\eta)}.
\end{equation}
\end{subequations}
To determine the structure of the steady-state colloid concentration, we balance convection and diffusion:
\begin{equation}
    \frac{\partial}{\partial \mathscr R}\left(D_N\frac{\partial N}{\partial \mathscr R} - V_\text{DP}N\right) = 0,
\end{equation}
where the diffusiophoretic velocity is given by equation~(\ref{eqn:Vdp_bruss}). Requiring a bounded solution, and defining the Péclet number as in equation~(\ref{eqn:peclet dimensional}), we get equation~(\ref{eqn:N_analytic}), which we compare to simulations in figure~\ref{hexagonsStripes}. We note that, depending on $f$, $\alpha$ can be positive or negative. In conjunction with the signs and magnitudes of $M_1$ and $M_2$, this property can flip the direction of $V_\text{DP}$. In the case of stripes, the sign change is trivially a phase shift. However, in the case of hexagons, our analytic model cannot accurately predict the length scale of colloids which are repulsed from the center, owing to complicated interactions with neighboring hexagons. In this case, one must rely on direct computation. For the same reason, our analysis cannot accurately predict any $\lambda_N > \lambda_C$. Finally, our model can only tell us $\lambda_N$ and not the maximum value of $N$; this value is sensitively dependent on early-time dynamics.

\subsection*{Numerical simulations}

Our numerical calculations were implemented with the open-source partial differential equations solver Basilisk~\cite{basiliskSource}. We modified their multi-grid solver of reaction-diffusion equations for the Brusselator model, which solves for $C_1$ and $C_2$, by including a diffusive tracer that is coupled in its advective term to $\nabla C_1$ and $\nabla C_2$. Basilisk's adaptive grid feature was employed in order to retain accuracy at low colloid length scales; we required the solver to refine the grid to maintain an absolute accuracy of 0.1 in $N$ (typically $\gg 1$ at the local maxima). The minimum and maximum refinement levels were set to 5 and 12, respectively. The time steps were elected by the solver. In each of our simulations, we analyzed only the steady state results; this was determined by inspection of simulation videos and time series of the domain maximum of $N$.

For all Brusselator simulations, we set $\ell_0=32\sqrt{k/D_{c_1}}$ and $\textrm{Da}_N=1$ (except in figure~\ref{Da} we vary $\textrm{Da}_N$). For the hexagons (figures~\ref{fish}, \ref{hexagonsStripes}(a), and~\ref{masterCurve}), we set $\mu=0.05$, $A=1.5$, and $D=4$. For the stripes (figures~\ref{fish},~\ref{hexagonsStripes}(b), and~\ref{masterCurve}), we set $\mu=0.04$, $A=2$, and $D=3$. By keeping $\mu$ small, we observed chemical wavelengths approximately equal to their critical value, simplifying our analysis. The initial conditions are spatially homogeneous with $C_1=A$ and $C_2=\sqrt{1+A\eta}(1+\mu)/A+[\text{noise}]$ where the noise is uniformly sampled in between -0.01 and 0.01. This choice of initial condition for $C_2$ represents deviations of $\mu$ above the critical concentration of $B$, with the noise perturbing the initial homogeneity. The boundary conditions are no-flux. For the colloid, the initial condition is spatially homogeneous with $N=1$ and no-flux boundary conditions are employed. $D_N$, $M_1$, and $M_2$ are varied systematically.

For the simulations corresponding to both \textit{Aracana Ornata} and \textit{Muraena lentiginosa}, we set the particle rates of diffusivity on the order of $10^{-4}$ relative to the solute, typical for micron-scale particles~\cite{alessio2021diffusiophoresis}. To achieve control over the pattern thickness, we vary the diffusiophoretic mobilities relative to the solute rate of diffusivity between $2.5\times10^{-2}$ and $10^{-1}$. Although this is smaller than typical experimental values, to our knowledge no experiments have been attempted to measure particle diffusiophoretic mobility in a biological membrane. Intuitively, we expect this setting would permit diffusiophoretic mobilities much lower than those observed in microfluidic experiments. We also expect that the shape anisotropy of chromatophores reduces their diffusiophoretic mobility. Furthermore, we realize that colloidal dense packing effects might play an important role in these systems owing to the steady state chemical gradient; such effects would reduce the diffusiophoretic convection.

We refer the reader to~\cite{nakamasu2009interactions} for the details of the cell-cell interaction model. We implemented this model using the reaction-diffusion module of Basilisk, utilized the exact same parameters as~\cite{nakamasu2009interactions}, and imposed a diffusiophoretic advection (using the advective tracer module of Basilisk) of low-diffusivity substances (chromatophores $u$ and $v$) in response to the gradient field of the high-diffusivity substance ($w$) with varying Péclet number. The two diffusiophoretic chromatophores are assigned equal and opposite mobilities. In figure~\ref{otherModels}, we employ 2D sinusoidal initial conditions with the observed wavelength and an overriding white noise field in the center, mimicking the pattern-forcing technique from~\cite{nakamasu2009interactions}. The same time step, box size, and adaptive grid technique as the Brusselator model implementation was applied. In figure~\ref{masterCurve}, we impose purely 2D sinusoidal initial conditions of a stable wavelength to simplify our length scale analysis.

We refer the reader to~\cite{yochelis2008formation} for a thorough implementation of the Gierer-Meinhardt model~\cite{gierer1972theory}. Using the reaction-diffusion module of Basilisk, we implemented this model with the same parameters and initial conditions as~\cite{yochelis2008formation}, electing $S=0.55$, and used Basilisk's advective tracer module to calculate the diffusiophoretic response of a third, non-reactive substance of varying Péclet number. The same time step, box size, and adaptive grid technique as the Brusselator model implementation was applied.

Post-processing was performed in Python 3.9, utilizing \textsc{numpy}, \textsc{scipy}, and \textsc{matplotlib}. The irregularly spaced output data was interpolated onto a 0.04$\times$0.04 square grid using \textsc{griddata} from the \textsc{scipy.interpolate} package. Custom colormaps were constructed using \textsc{linearsegmentedcolormap} from the \textsc{matplotlib.colors} package. Calculations of $\lambda_N$ were performed by identifying the locations (in two dimensions for hexagons and in the one-dimensional average for striped regions) of the maxima in colloid concentration using the \textsc{peak\_local\_max} function from the \textsc{skimage.feature} package, histogramming the data points to obtain relative colloid concentration as a function of radial distance from the maximum, and identifying the exponential decay distance for the average curve. Error was calculated by identifying the decay distances after adding and subtracting one standard deviation to the average curve.

\section*{Data availability}
Data supporting the plots in this paper are available from the corresponding author upon reasonable request.

\section*{Code availability}
The codes used for the Basilisk simulations in this study can be reproduced by modifying the Basilisk source Brusselator example at \url{http://basilisk.fr/src/examples/brusselator.c} and are available on github: \url{https://github.com/balessio/diffusiophoresis_turingPatterns.git}.

\section*{Acknowledgments}
We thank the Birch Aquarium at the Scripps Institution of Oceanography for providing high-resolution photographs of \textit{Aracana ornata}. A.G. thanks the National Science Foundation (CBET - 2238412) CAREER award for financial support. Acknowledgement is made to the donors of the American Chemical Society Petroleum Research Fund for partial support of this research. We thank Anirudha Banerjee, Ritu Raj, Juan Santiago, Dan Schwartz, Howard Stone, Kaare Jensen, and Ananth Govind Rajan for their helpful feedback on the manuscript.

\section*{Author contributions}
B.M.A. and A.G. conceived the project. B.M.A. designed the simulations with Basilisk and processed the results. B.M.A. and A.G. conducted the theoretical analysis. B.M.A. and A.G. discussed the results and wrote the paper. 

\section*{Competing interests}
The authors declare no competing interests.

\section*{Supplementary Information}

\setcounter{figure}{0}
\renewcommand{\thefigure}{S\arabic{figure}}

\begin{figure*}[t]
\centering
\includegraphics[width=0.95\textwidth]{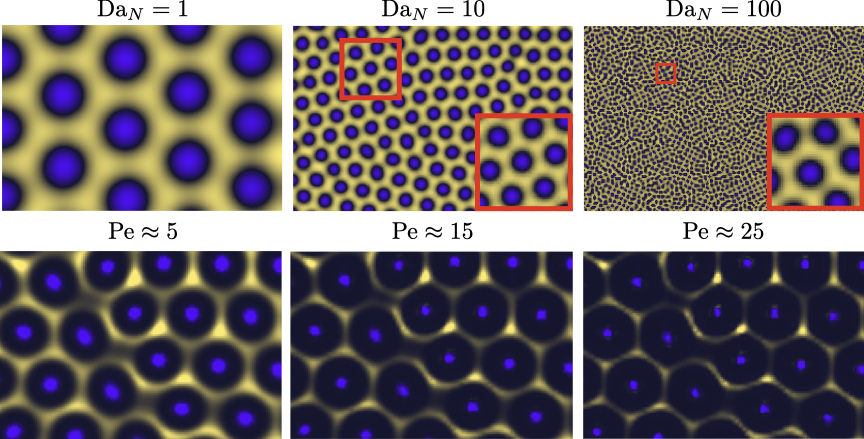}
\caption{\textbf{Size separation of $\ell_\textrm{p}$ and $\ell_{\textrm{t}}$}. (Top row) A two-component Brusselator model is unable to change the size and the thickness independently. The simulations are done with different $\textrm{Da}_N$ but without any convection terms such that $\textrm{Pe}$=0. (Bottom row) Upon inclusion of diffusiophoresis, we are able to control the edge-thickness of the hexagons by changing the $\textrm{Pe}$ while keeping the size of the hexagons constant.}\label{Da}
\end{figure*}

\begin{figure*}[t]
\centering
\includegraphics[width=0.95\textwidth]{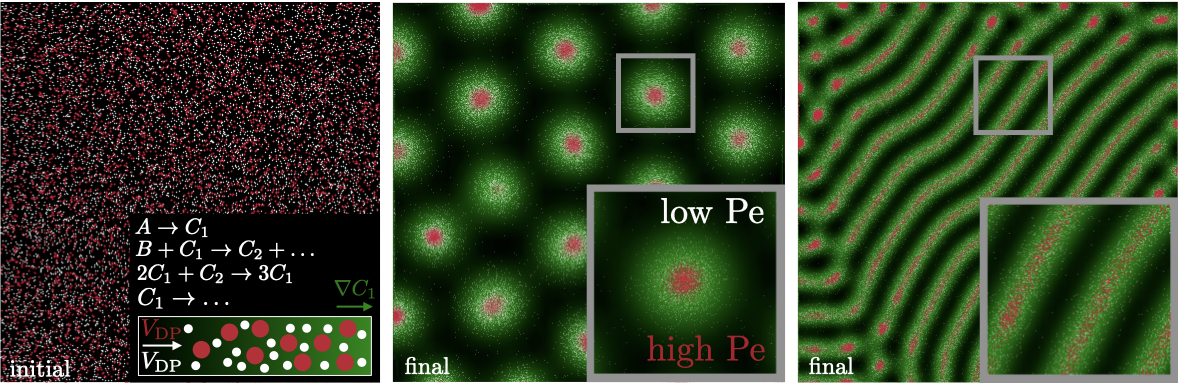}
\caption{\textbf{Schematic of diffusiophoresis-enhanced Turing patterns}. (Left) At $t=0$, the Brusselator model creates reaction-diffusion gradients in the solute concentrations $C_1$ and $C_2$ that induce motion via diffusiophoresis in the two colloidal particles, denoted by white (low P\'{e}clet) and red (high P\'{e}clet) circles. The simulations yield hexagon (middle) and stripe (right) patterns. The colloids' characteristic length scale is controlled by their Péclet number, whereas the size of the hexagon and the distance between the stripes is dictated by the solute concentration gradient, denoted by the green halo above.}\label{schematic}
\end{figure*}

\section*{Supplementary Video 1}
Diffusiophoresis-enhanced chemical Turing pattern simulated with the Brusselator model. Four species are present: blue and yellow chromatophores and two molecular solutes (one shown on left; yellow to blue color scheme represents increasing concentration). The nonlinear coupled reaction and diffusion of the two solutes produces a diffuse form of the hexagon pattern. The two chromatophores organize by a convective-diffusive process, where they convect along the gradients of the solutes and settle along local extrema in solute concentration. The yellow species has a diffusiophoretic velocity opposite in sign to the blue species.

\section*{Supplementary Video 2}
Diffusiophoresis-enhanced chemical Turing pattern simulated with the Brusselator model. Four species are present: dark-brown and light-tan chromatophores and two molecular solutes (one shown on left; tan to light-tan color scheme represents increasing concentration). The nonlinear coupled reaction and diffusion of the two solutes produces a diffuse form of the hexagon pattern. The two chromatophores organize by a convective-diffusive process, where they convect along the gradients of the solutes and settle along local maxima in solute concentration. The light-tan species has a lower rate of diffusivity than the dark-brown species, so it concentrates within a shorter radius from the maxima, forming a “double spot” pattern.

\section*{Supplementary Video 3}
Diffusiophoresis-enhanced chemical Turing pattern simulated with the cell-cell interaction model. Three species are present: blue and yellow chromatophores (right) and a molecular solute (left; yellow to blue color scheme represents increasing concentration). In this model, these three species are coupled in their reactions. This effect, combined with diffusion, produces intricate patterns. The two chromatophore species furthermore convect diffusiophoretically along the gradients of the molecular solute, organizing separately along the maxima (blue) and the minima (yellow) of the solute concentration.

\section*{Supplementary Video 4}
Diffusiophoresis-enhanced chemical Turing pattern simulated with the Gierer-Meinhardt model. Four species are present: dark-brown and light-tan chromatophores (right) and two molecular solutes (one shown on left; tan to light-tan color scheme represents increasing concentration). The nonlinear coupled reaction and diffusion of the two solutes produces a diffuse form of the pattern. The two chromatophores organize by a convective-diffusive process, where they convect along the gradients of the solutes and settle along local maxima in solute concentration. The light-tan species has a lower rate of diffusivity than the dark-brown species, so it has a sharper pattern.






\end{document}